# Investigate a dense transparent glass doped $Nd^{3+}$ ion for high power laser applications

**Research Project by:**

A A El-Maaref, Essam Abdel Wahab, and Kh Shaaban

Al-Azhar university, Assiut, Egypt

**Abstract**

Silicate glasses system with different chemical formula as $(50-x) SiO_2 - 50 Bi_2O_3 - xNd_2O_3$ will be prepared by melt quenching method. Using X-ray diffraction, the amorphous nature of these glasses will be identified. The structure changes of these glasses will be studied by FT-IR spectra. The mechanical properties will be associated with the results of FT-IR spectra. Optical characteristic will be measured and investigated at ordinary temperatures (room temperature). Polarizability and basicity of glasses according to the refractive index and optical band gap will be investigated. Judd–Ofelt theory will be applied to analyze the spectra arising from $Nd^{3+}$ doped bismuth silicate glasses. The intensity parameters $\Omega_{2,4,6}$ of the present complex and lifetimes of selected levels will be theoretically calculated, as well. Also, the possibility of getting laser transitions between different levels of $Nd^{3+}$ doped present glasses will be investigated by calculating the lifetime and branching ratio of the transitions among these levels. Emission and absorption cross sections of the prepared materials will be measured at different spectral regions.

**Introduction**

Glass is an amorphous substance and has a considerable potential in different fields [1-5]. Glass materials are preferred in the advanced applications such as laser, medical, or military uses. Transparency and durability are of interest as prominent features of glass properties [1-10]. Silicate, borate, phosphate, germinate, arsenate etc. are all different types of glass. Glasses of heavy metal oxide (HMO), that's glasses include about 50 mol of a heavy metal cation percent like $TeO_2$, $Sb_2O_3$, $Bi_2O_3$, $Ga_2O_3$, and PbO [11].

These glasses have photon energy smaller than other glasses, and higher refractive index. Therefore, the HMO glasses are transparent, promising materials for optoelectronics, fibers, sensors, optical instruments, and offer superior optical, electrical, and radiation shielding materials properties [12-15]. Glasses can be formed from $SiO_2-Bi_2O_3$, to increase its durability, another oxide should be added. Neodymium, with the symbol *Na* and Atomic No. 60, is the first element of lanthanide series. $Nd_2O_3$ improve the chemical resistance of the glass matrix, and its optical characteristics. Nowadays, one of the most important research areas focuses on the development of rare-earth (RE)-doped glasses for use in communication systems, such as fibers, optical amplifiers, and lasers. The present work aims to report the structural, optical, mechanical properties of a newly synthesized glass system with the composition(50-x)

$SiO_2 - 50\ Bi_2O_3 - xNd_2O_3$ and use of these glasses as conductive materials for energy applications.

**Previous Studies**

Recently, great efforts have been performed to study the structure and physical properties of $Nd^{3+}$ doped different materials. Ramteke et al. [16] have studied the effect of $Nd^{3+}$ on spectroscopic properties of lithium borate glasses. El-Maaref et al. [7] have studied the optical properties and radiative rates of $Nd^{3+}$ doped zinc-sodium phosphate glasses. UV-vis spectra of these glasses were analyzed at different concentrations of $Nd_2O_3$. The effect of neodymium concentration on the density and energy band gap has been investigated. The density of the present glasses slightly increases with the increasing of $Nd_2O_3$. Judd-Ofelt theory was used to determine the optical parameters such as line strengths, optical intensity parameters ($\Omega^t$), transition probabilities, and transition lifetimes. Hypersensitive transitions were identified in the absorption spectrum, the greatest line strengths are recorded at the transitions $2G_{7/2}$ - $4G_{5/2}$, $4S_{3/2}$ - $4F_{7/2}$ and $4D_{1/2}$ - $4D_{3/2}$ - $4D_{5/2}$ - $2I_{11/2}$ with wavelengths of 580, 475 and 355 nm, respectively. Lifetimes of the important $4F_{3/2}$ laser-level were determined; which show decreasing trend with the increasing of $Nd_2O_3$ content and are found to be between 0.838 and 1.595 ms [17-20].

Neodymium doped lithium lead alumino borate glasses has been synthesized with the chemical composition $10Li_2O-10PbO-(10-x)\ Al_2O_3-70B_2O_3-x\ Nd_2O_3$. The present material exhibit tow emission lines at 1063 nm and 1350 nm, these lines originated from $^4F_{3/2}$ - $^4I_{11/2}$ and $^4F_{3/2}$ - $^4I_{9/2}$, respectively [21]. Trejgis et al. [22] investigated the suitability of $Nd^{3+}$-doped oxyfluorotellurite $(65-x)TeO_2–20ZnF_2–12PbO–3Nb_2O_5–xNd_2O_3$ (x = 0.1, 1, 2, 5 and 10) glasses (TZPN) for SBR-based luminescent thermometry. Xu et al. [23] have investigated the effect of $Nd^{3+}$ impurity in the phosphate laser glasses.

Kolobkova et al. [24] have studied the spectroscopic properties and energy transfer of fluorophosphate glasses doped with $Nd^{3+}$, Judd-Ofelt theory has been used to evaluate optical intensity parameters, radiative lifetimes, and emission cross sections of $Nd^{3+}$ doped fluorophosphate. The emission bands at NIR and IR spectral regions $Nd^{3+}$ ions doped zinc-lithium fluoroborate glasses have been studied and analyzed [25]. Djamal et al. [26] have studied the spectroscopic properties of $Nd^{3+}$ ion-doped Zn-Al-Ba borate glasses for near infra-red emitting device applications. Lead fluorosilicate glasses doped $Nd^{3+}$ for photonic device applications have been prepared and optically characterized by Manasa et al. [27]. Structural and optical properties of borosilicate glasses doped $Nd^{3+}$ have been experimentally analyzed using XRD, FTIR, UV-Vis spectroscopy, and excitation emission spectroscopy. As well as, the optical intensity parameters and radiative rated of these glasses have been computed using Judd-Ofelt theory [28]. In other work [29], Judd-Ofelt theory has been applied to predict the optical parameters and radiative rates of $Nd^{3+}$-doped multicomponent tellurite glasses.

**Research methodology**

1. SBN glasses with the chemical formula (50-x) $SiO_2$ – 50 $Bi_2O_3$ – $xNd_2O_3$, will be fabricated by the melting quenching method at 1200 °C in ceramic containers for 3 hours and annealed at 450 °C for 1 hour.
2. The states of the obtained glasses will be identified by X-ray diffractometer.
3. The structure changes of these glasses will be studied using Bruker's VERTEX 70 FTIR Spectrometers.
4. The mechanical measurements will be determined using a pulse-echo technique (KARLDEUTSCH Echograph model 1085).
5. The optical properties such as refractive index and energy band gap will be studied by spectrophotometer (typeJASCO V- 670) in the wavelength range from 200 to 2500 nm.
6. The experiential setup for measuring the emission spectra consists of a diode-pumped CW laser with emission wavelength of 355 nm as a pumping source, laser beam reshaping optics for collimating the laser beam, infinitely corrected objective and a UV–Vis spectrometer. The power of the diode-pumped CW laser is controlled by two means, first through the adjustment of the power of the pumping diode and second through a round variable neutral density filter. The laser beam passes through a 50/50 beam splitter to the sample and a power meter. The pumping laser spot is expanded collimation by a set of plano-concave and plano-convex cylindrical lenses.